\documentclass[twocolumn,showpacs,preprintnumbers,amsmath,amssymb]{revtex4}
\usepackage{graphicx}% Include figure files
\usepackage{dcolumn}% Align table columns on decimal point
\usepackage{bm}% bold math

\newcommand{\bi}{\bigskip}
\newcommand{\no}{\noindent}
\newcommand{\be}{\begin{eqnarray}}
\newcommand{\ee}{\end{eqnarray}}
\newcommand{\hk}{\hspace{0.1cm}}

\newcommand{\rk}{\right)}
\newcommand{\lk}{\left(}

\newcommand{\hbo}{\hbox to 1 true cm {\hfill } } 

\begin{document}

\preprint{}

\title{Center flux correlation in SU(2) Yang-Mills theory } 
% Force line breaks with \\

\author{K. Langfeld, G. Schulze and H. Reinhardt}
 \affiliation{Institut f\"ur Theoretische Physik, 
Auf der Morgenstelle 14\\
D-72076 T\"ubingen, 
Germany}%Lines break automatically or can be forced with \\
%\affiliation{%

%\affiliation{

\date{August 3, 2005}% It is always \today, today,

\begin{abstract}
By using the method of center projection the center vortex part of the gauge
field is isolated and its propagator is evaluated in the center Landau
gauge, which minimizes the open 3-dimensional Dirac volumes of non-trivial
center links bounded by the closed 2-dimensional center vortex surfaces. The
center field propagator is found to dominate the gluon propagator (in Landau
gauge) in the low momentum regime and to give rise to an
OPE correction to the latter of ${\sqrt{\sigma}}/{p^3}$.The screening mass of
the center vortex field vanishes above the critical temperature of the
deconfinement phase transition, which naturally explains the second order 
nature of this transition consistent with the vortex picture. Finally,
the ghost propagator of maximal center gauge is found to be infrared
finite and thus shows that the coset fields play no role for confinement.
\end{abstract}

\pacs{11.15.Ha, 12.38.Aw, 12.38.Gc }
\maketitle

\no
One of the fundamental problems of particle physics is the understanding of
confinement of quarks and gluons in QCD. Although confinement has not yet been
thoroughly understood, several pictures of confinement have been developed, 
which received strong support by lattice calculations in recent years. 
Among these are the dual Meissner effect and the center vortex 
condensation (for a recent review see~\cite{Greensite:2003bk}). 
In particular, the center vortex picture is quite appealing:
center vortices identified by the method of center 
projection~\cite{DelDebbio:1996mh} are physical in the sense of showing
the proper scaling~\cite{Langfeld:1997jx}. Moreover, the 
deconfinement phase transition at finite temperature emerges 
as depercolation 
transition~\cite{Langfeld:1998cz,Engelhardt:1999fd,Langfeld:2003zi}.
Center vortices also provide a geometric interpretation of topological 
charge in terms of  intersection and writhing of vortex surfaces or 
loops~\cite{Engelhardt:1999xw,Reinhardt:2001kf}. 
\bi

\no
A different confinement mechanism was proposed by Gribov~\cite{Gribov:1977wm} 
and further elaborated  by Zwanziger~\cite{Zwanziger:1991ac}. 
This mechanism is based on the infrared dominance of the field 
configurations near the Gribov horizon, which gives rise to an infrared 
singular ghost propagator, which  is considered to be a signal
of confinement~\cite{Alkofer:2000wg}. 
In Landau gauge this infrared singularity disappears, when center vortices 
are eliminated from the Yang-Mills ensemble~\cite{Gattnar:2004bf}. 
Since the infrared singularities are caused by field configurations on the
Gribov horizon, one expects, that center vortices are on the Gribov horizon,
which indeed can be shown to be the case~\cite{Greensite:2004ur}. 
The results of~\cite{Gattnar:2004bf} and~\cite{Greensite:2004ur} show 
that, in Landau and Coulomb gauge, the center vortices are not only on 
the Gribov horizon, but they also dominate the infrared physics. This 
suggests, that the center vortices may be the confiner in any gauge, which is
very plausible since center vortices can, in principle, be defined in a gauge
invariant way and after all confinement is a gauge independent phenomenon. 
\bi

\no
In this paper we will further elaborate on the connection of the two 
confinement scenarios described above, i.e. on the interplay between 
center vortices and ghosts. Using the method of center 
projection~\cite{DelDebbio:1996mh}, we separate the confining
center degrees of freedom from the remaining (non-confining) coset degrees of
freedom and calculate the associate propagators as well as the corresponding
ghost propagator. We will find, that in 
contrast to the familiar Landau gauge, in the maximum center gauge the ghost
propagator is infrared finite and thus shows no signal of confinement. This
result is not surprising since in this gauge the Faddeev-Popov operator 
does not feel the center part of the gauge field. The confining center 
degrees of freedom constitute the center vortex field. 
We calculate its propagator which is carefully extrapolated to the 
continuum limit. We will find, that the center field propagator
dominates the infrared behavior of the gluon propagator, while the 
ultraviolet behavior of the latter is exclusively determined by the coset 
field propagator. However, the center vortex field gives rise to an 
correction to the gluon propagator of the form
$\sqrt{\sigma}/p^3$, which indicates its relevance in the context 
of the operator product expansion. 
\bi

\no
Although the lattice provides a gauge invariant approach to Yang-Mills theory
the transition to the continuum theory is facilitated by using a gauge, in
which the fields are smooth. The prototype of such a gauge is the well known
Landau gauge
\be
\label{1}
\sum_{x, \mu} tr U^\Omega_\mu (x) \stackrel{\Omega }{\longrightarrow } 
\mathrm{max} \hk ,
\ee
where $U^\Omega_\mu (x) = \Omega (x) U_\mu (x) \Omega (x + \hat{\mu})$ is the
gauge transform of the link variable $U_\mu (x)$. This gauge brings the 
links as
close as possible to the unit element of the gauge group and one therefore
expects, that the gauge fields $U_\mu (x)$ will be smooth in this gauge. For
smooth links near the unity we can extract the continuum gauge 
field $A_\mu (x)$ defined by ($a \to 0$, $a$ the lattice spacing) 
\be
\label{2}
U_\mu (x) = e^{i\,  a A_\mu (x)}
\ee
in the standard fashion by Taylor expansion. In this (naive) continuum limit 
the gauge condition (\ref{1}) 
reduces to the usual (continuum) Landau gauge $\partial_\mu A_\mu = 0$. 
In this
paper we will use various modifications of the Landau gauge to identify the
center vortex content and the remaining coset part of the gauge field as 
will be detailed further below. 
\bi

\begin{figure}
\includegraphics[height=7cm]{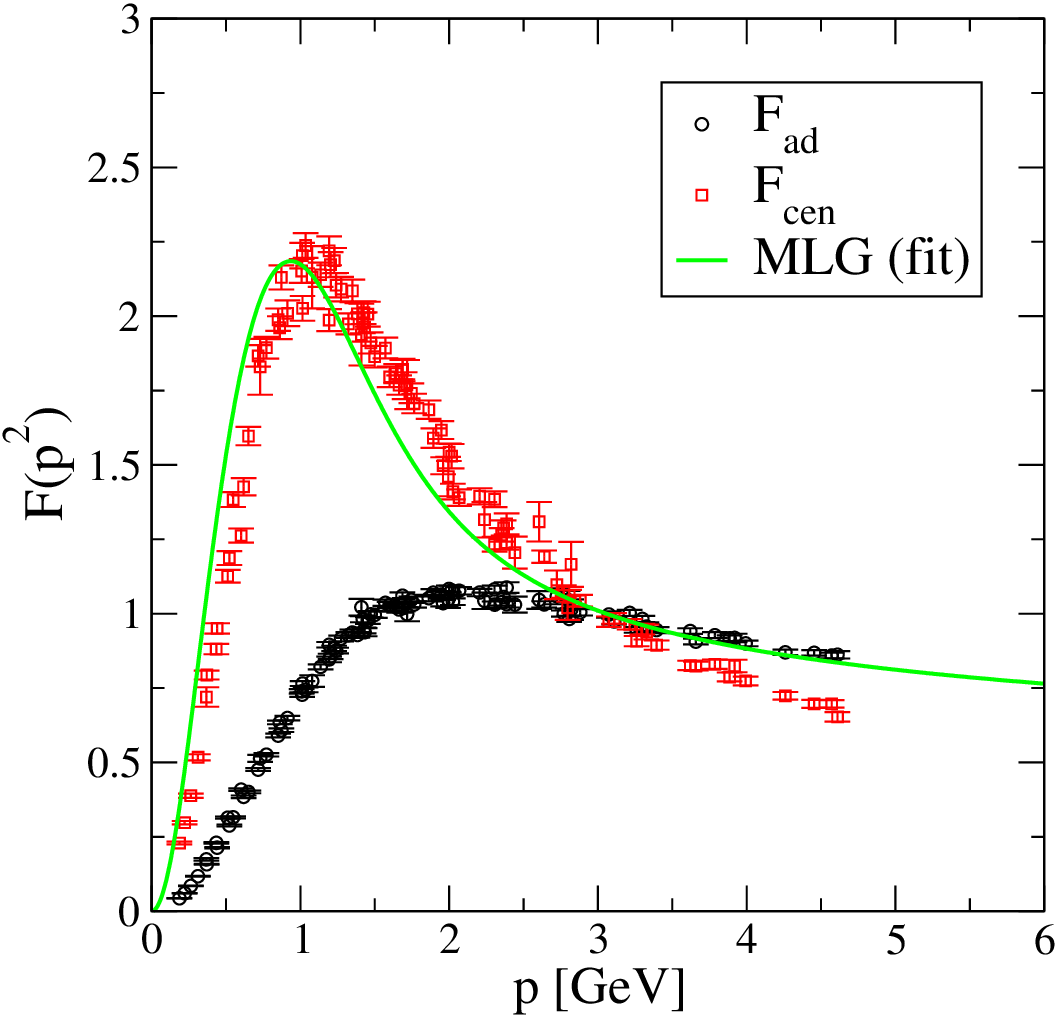} %\hspace{0.5cm}
\caption{\label{fig:fgluon} The gluon formfactor $F_\mathrm{ad}(p^2)$ 
in MCG as function of the momentum transfer $p$ (circles), the center 
field form factor (squares) and the fit to the  gluon form factor 
in Minimal Landau gauge (line). Normalization: $F(p=3 \, \mathrm{GeV})=1$. 
}
\end{figure}
\no
To identify the center vortex content of a gauge field, we use the method of
center projection~\cite{DelDebbio:1996mh}, 
which is based on the so-called maximum center
gauge defined by 
\be
\label{3}
\sum_{x, \mu} \Bigl[ \mathrm{tr} U^\Omega_\mu (x) \Bigr]^2 
\stackrel{\Omega }{\longrightarrow } 
\mathrm{max} \hk ,
\ee
This gauge fixes the gauge group only up to center gauge transformations, i.e.
it
fixes only the coset $SU (2) / Z (2) = SO (3)$ and brings a given link
$U_\mu (x)$ as close as possible to a center element ($\pm1$ for $SU (2))$. 
Once, this gauge is
implemented, center projection implies to replace a link $U_\mu (x)$ by its
closest center element, which is given by 
\be
\label{4}
Z_\mu (x) = \mathrm{sign} \; \mathrm{tr} \,  U^\Omega_\mu (x) \hk .
\ee
The center projected configurations $Z_\mu (x)$ form 3-dimensional volumes of
links $Z_\mu (x) = - 1$, the closed boundaries of which represent the center
vortices. These vortex surfaces are formed by dual plaquettes 
$P_{\mu \nu} (x) = - 1$.
We separate the center projected vortices from
the original gauge fields, by writing~\cite{deForcrand:1999ms} 
\be
\label{2A*}
U^\Omega_\mu (x) = Z_\mu (x) \bar{U}_\mu (x) \hk .
\ee
\bi

\begin{figure}
\includegraphics[height=7cm]{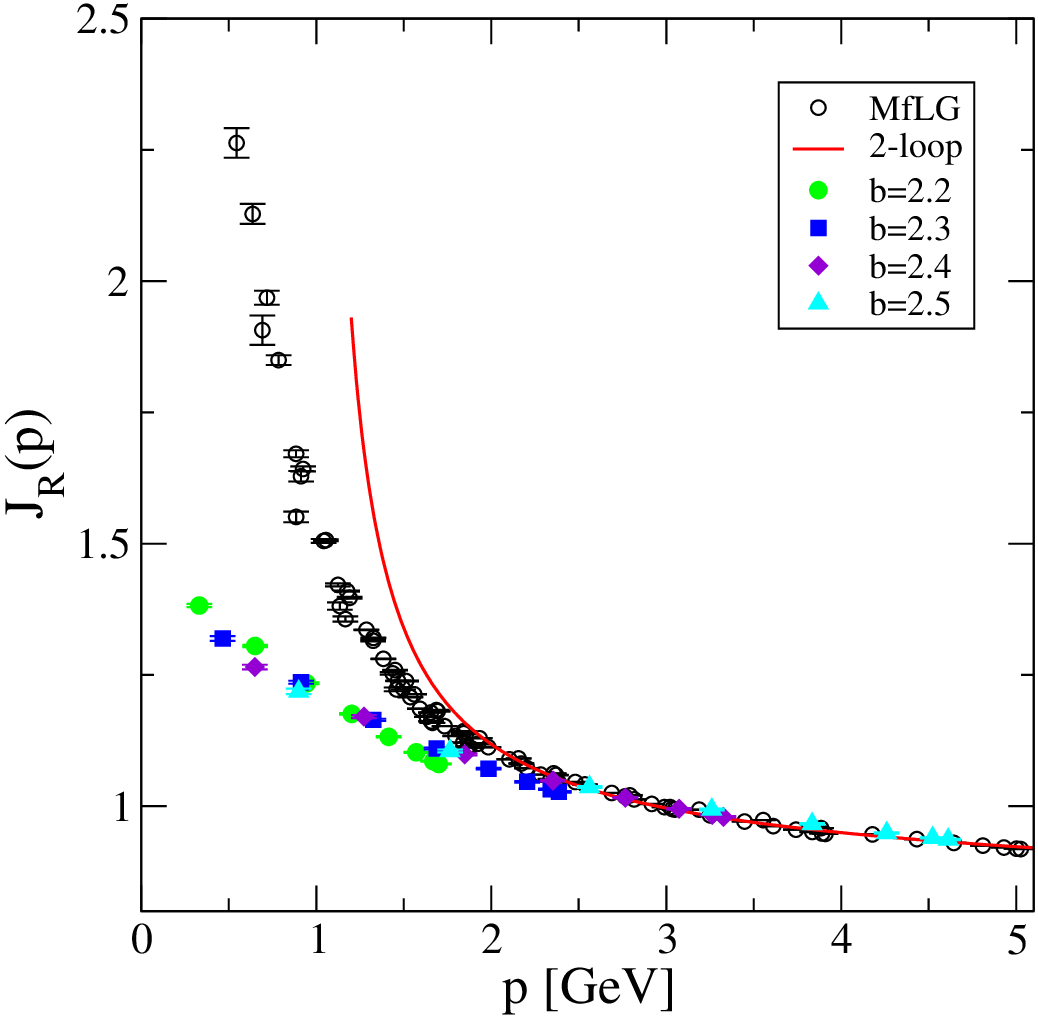} %\hspace{0.5cm}
\caption{\label{fig:ghost} The ghost form factor (full symbols)
   in MCG as function of momentum transfer. The ghost form 
   factor in Minimal Landau gauge (open symbols, data 
   from~\cite{Bloch:2003sk}). Two loop  perturbation theory (solid line). 
} 
\end{figure}
\no
The maximal center gauge is known to be just the minimal Landau gauge for the
adjoint representation, which does not feel the center. 
Thus, the links $\bar{U}_\mu (x)$, defined in the MCG, should be 
sufficiently closed to $1$ and we can extract a continuum field
$\bar{A}_\mu (x)$ in the standard fashion $\bar{U}_\mu (x) = \exp \lk i\,  
a \bar{A}_\mu (x) \rk \hk $. With the parameterization $U^\Omega_\mu (x) = 
a^0_\mu (x) + i \vec{a}_\mu (x) \vec{\tau} $, we find
$ a \bar{A}^b_\mu (x) \; = \; 2 \, a^0_\mu (x) \, a_\mu (x) ^b $. 
In view of (\ref{2A*}) we interprete the $\bar{A}_\mu (x)$ as the gluonic
radiative fluctuations around the center vortex ``background''. 
\bi

\no
Since
$\bar{U}_\mu$ satisfies the adjoint Landau gauge, the radiation field
$\bar{A}_\mu (x)$ is transversal and its propagator can be expressed as
\be
\Bigl\langle \bar{A}_\mu (p) \bar{A}_\nu (- p) \Bigr\rangle = 
\lk \delta_{\mu \nu} - \frac{p_\mu
p_\nu}{p^2} \rk \frac{F _\mathrm{ad} (p^2)}{p^2}  \hk ,
\ee
where $F_\mathrm{ad} (p^2)$ is the adjoint gluon form factor.  

\vskip 0.3cm 
We have performed a large scale study of SU(2) lattice gauge theory 
using $16^4$ and $24^4$ lattices and $\beta $ values of the Wilson action 
in the range $\beta \in [2.15, 2.5]$.  
The result for the (renormalized) adjoint gluon form factor is shown in 
figure~\ref{fig:fgluon} (details of the numerical approach will 
be published elsewhere). For large momenta this quantity reproduces the
perturbative result for the full gluon form factor in Landau gauge. 
Like the full gluon form factor, $F_\mathrm{ad} (p^2)$ vanishes 
for $p \to 0$, signaling a mass gap in the excitation spectrum of 
$\bar{A}_\mu (x)$. Finally, $F_\mathrm{ad} (p^2)$ deviates essentially 
from the  gluon form factor of Minimal Landau Gauge in the
intermediate momentum regime.
\bi

\no
It has been well established by now, that for SU (2) the center projected
vortices, defined by the $Z_\mu (x)$ (\ref{4}), basically produce the full
string tension, while the radiative coset field $\bar{U}_\mu$ 
(or $\bar{A}_\mu$)
does not contribute to confinement. Since, the MCG condition (\ref{3}) depends,
only on the coset field $\bar{U}_\mu$, we expect the corresponding ghost
propagator to show no signal of confinement. 
The Faddeev-Popov operator of the MCG is given by
\be 
M^{ab}(x,y) &=& \sum _{\mu } \biggl\{ 
- \, \left[ f^{ab}(x-\mu) + f^{ab}(x) \right] \; \delta (x,y) 
\nonumber \\ 
&+& \left[ f^{ab}(x-\mu) - g^{ab}(x-\mu) \right] \; \delta (x-\mu ,y) 
\nonumber \\ 
&+& \left[ f^{ab}(x) + g^{ab}(x) \right] \; \delta (x+\mu ,y) 
\biggr\} \; , 
\label{eq:fad_ex} 
\ee 
where
\be 
f^{ab}(x) &=& \left(a_\mu ^0 (x) \right)^2 \delta ^{ab} - a^a_\mu (x)  
a^b _\mu (x) \; , 
\nonumber \\ 
g^{ab}(x) &=& a_\mu ^0 (x) \, \epsilon ^{abc} \, a^c _\mu (x) \; .
\nonumber 
\ee
Figure~\ref{fig:ghost} shows the ghost form factor $J(p^2)$ of 
the MCG defined by
\be
\label{11}
\Bigl\langle M^{- 1} \Bigr\rangle ^{ab} (p) \; = \; \delta ^{ab} 
J (p^2) /p^2 \hk .
\ee
At high momenta it approaches the ghost form factor of the Minimal Landau Gauge
but differs drastically from the latter in the infrared: While the ghost 
form factor
of the Minimal Landau gauge is infrared divergent (what is considered as a
signal of confinement), the one of MCG, which does not feel the center,
seems to be infrared finite. This is consistent with the result obtained 
in~\cite{Gattnar:2004bf}, where 
it was shown, that the infrared divergent behavior of the ghost form
factor in minimal Landau gauge disappears, when center vortices are 
removed from the Yang-Mills ensemble. 
\bi

\begin{figure}
\includegraphics[height=7cm]{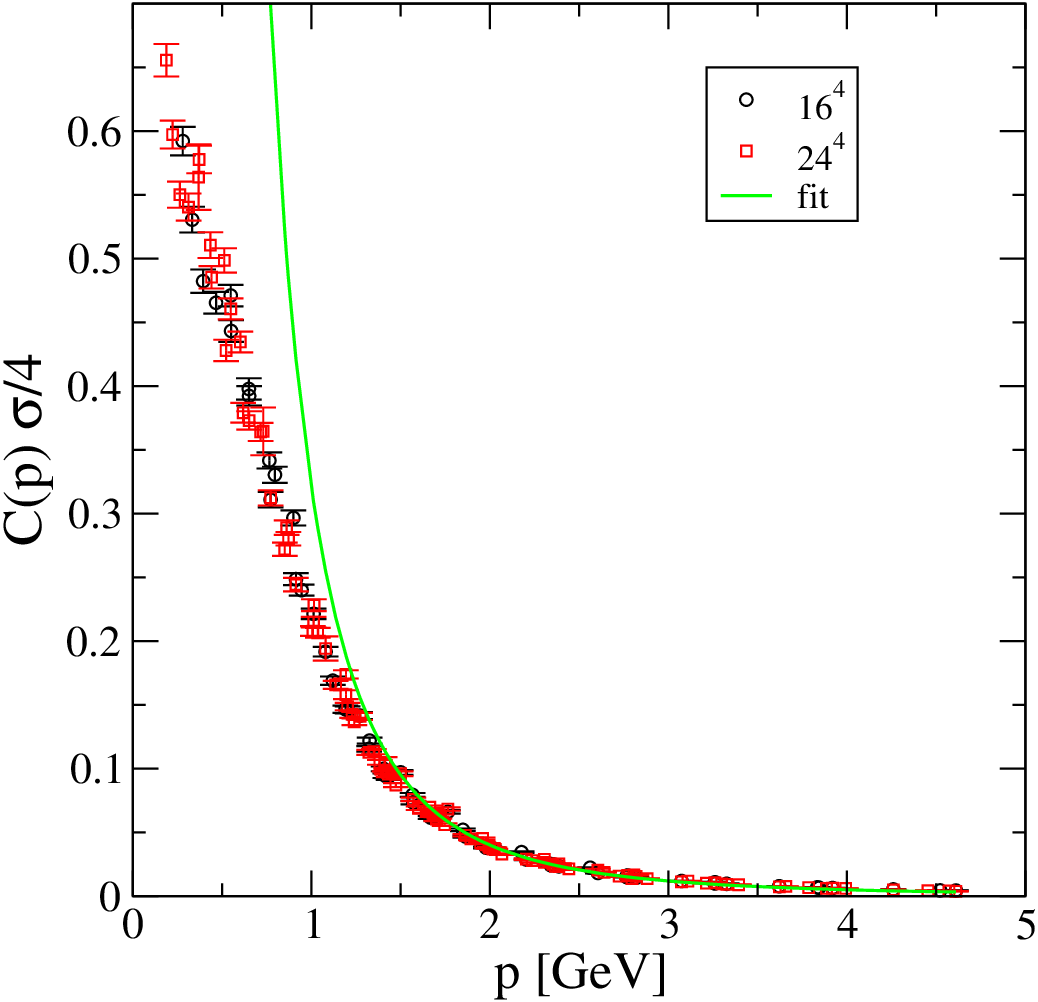} %\hspace{0.5cm}
\caption{\label{fig:cenp} Center field correlation function as function 
   of momentum for two different lattice sizes. 
}
\end{figure}
\no
We also use the analog of the Landau gauge for the center
projected fields 
\be
\label{12}
\sum_{x, \mu} Z_\mu (x) \stackrel{\Omega _2}{\longrightarrow } 
\mathrm{max} \,  \hk , 
\ee 
where $\Omega _2 $ are $Z_2$ gauge transformations. Each link $Z_\mu
(x)$ defines an elementary cube on the dual lattice and the total number of
non-trivial center links $Z_\mu (x) = - 1$ defines  open
hypersurfaces $\Sigma$ bounded by closed center vortex surfaces $\partial
\Sigma$. The
Landau center gauge condition, (\ref{12}), minimizes the 
number of $Z_\mu (x)
= - 1$ links and thus minimizes the volume of 
the open 3-dimensional hypersurfaces $\Sigma$. Since, the
minimal open hypersurfaces are completely determined by their
boundary $\partial \Sigma $, 
it is expected that they scale properly towards the continuum limit,
if the vortex surfaces $\partial \Sigma$ do. 
Let ${\cal P}$ be the probability that a given link element $Z_\mu (x)$ 
is negative. If the ratio between the minimal cubic volume and 
the space-time volume, i.e., 
$$ 
\frac{ {\cal P} \; 6 \, N^3 \, N_t \; a^3 }{ 
 N^3 \, N_t \; a^4 } \; = \; \frac{ 6  \, {\cal P} }{a} \; , 
$$
is a physical quantity in the continuum limit $a \rightarrow 0$, 
the probability must scale according ${\cal P} \; = \; \kappa \; a$, 
with $\kappa $ being independent of $a$. This is indeed confirmed by 
lattice calculations~\cite{Kovalenko:2004xm}.  
Our numerical calculations yield $\kappa \approx 0.26 (1) \sqrt{\sigma}$.
\bi

\begin{figure}
\includegraphics[height=7cm]{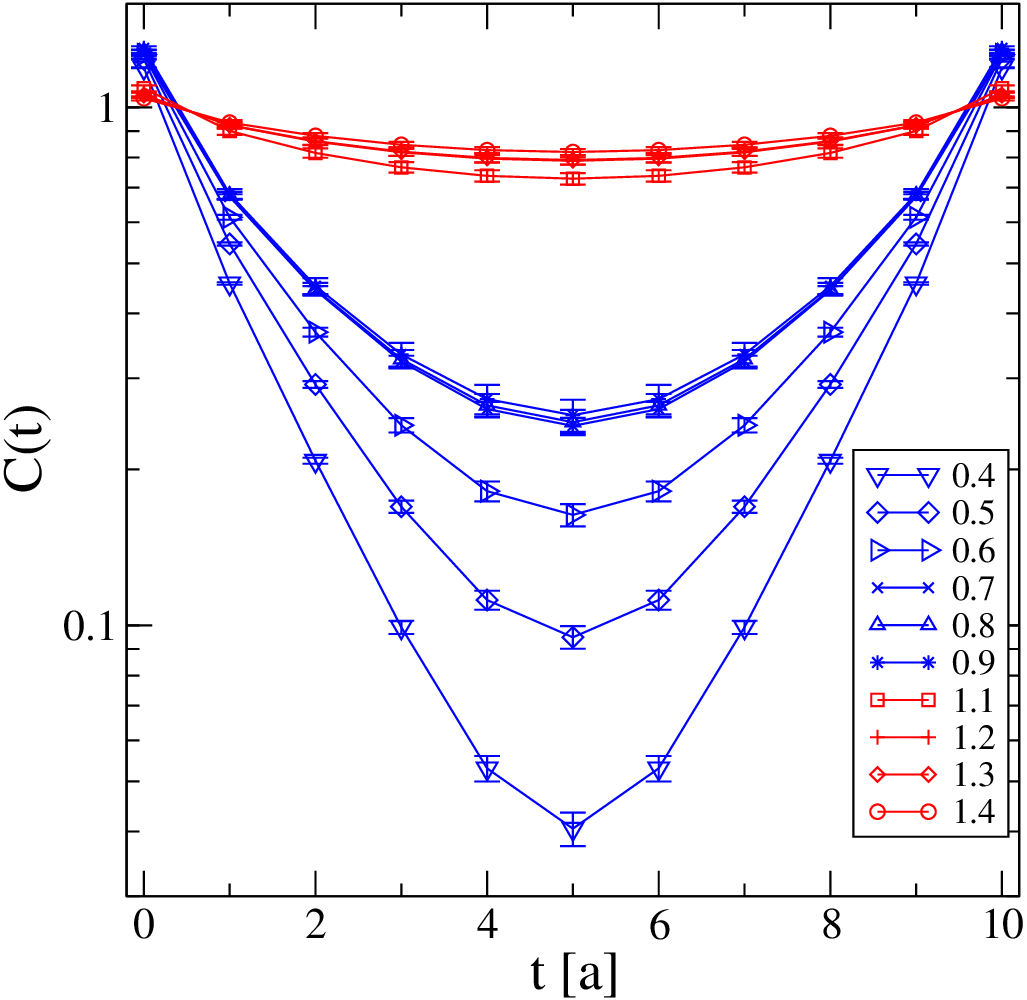} %\hspace{0.5cm}
\caption{\label{fig:ct} Center field correlation function as function  
  of Euclidean time for several temperatures (in units of the critical 
  temperature). 
}
\end{figure}
\no
Consider now the connected center field correlation function 
\be
\label{18}
a^2 c_{\mu \nu} (x-y) = \langle Z_\mu (x) Z_\nu (y) \rangle 
\; - \; \langle Z_\mu (x)
\rangle \langle Z_\nu (y) \rangle \hk . 
\ee 
The factor $a^2$ was introduced for later convenience. Although 
$Z_\mu (x)$ are integer valued fields, the function $c_{\mu \nu} (x-y)$, 
emerging from an ensemble average, appears to be smooth. 
Since the center Landau gauge condition (\ref{12}) is equivalent to 
\be
\label{25}
\sum_\mu \Bigl[ Z_\mu (x + \hat{\mu}) - Z_\mu (x) \Bigr] \;  = \; 0 \hk , 
\ee
the center field propagator is transverse: $ \partial _\mu \; 
c_{\mu \nu} (x-y) \; = \; 0 $. 
We therefore introduce the center field form factor $F_\mathrm{cen} (p)$ 
as usual by 
\be
C_{\mu \nu }(p) = \left( \delta _{\mu \nu } - 
\frac{p_\mu p_\nu }{p^2} \right) \; C(p^2) \; , \; 
C(p^2) = \frac{F_\mathrm{cen} (p)}{p^2} . 
\ee
\bi

\no
One of our important findings is that the 
center field correlation function $C_{\mu \nu }(p)$ 
is independent of the lattice spacing. The consequences are two-fold: 
Firstly, the propagator  $C_{\mu \nu }(p)$ is {\it not} subjected to wave 
function renormalization. Secondly, $C_{\mu \nu }(p)$ behaves as a genuine 
gluonic correlation function with mass dimension two. 

\vskip 0.3cm
Our numerical results for the propagator 
$ C_{\mu \nu }$ in momentum space are shown in figure~\ref{fig:cenp}. 
Most striking is that the high momentum tail is well fitted by the power 
law (see figure~\ref{fig:cenp}) 
\be 
C(p) /4 \; = \; \frac{ 3.7(1) \; \sqrt{\sigma } }{ p^3 } \; , 
\hbo p \ge 2 \, \mathrm{GeV} \; . 
\label{eq:fit} 
\ee 
This implies that the center field correlator is sub-leading in the 
high momentum regime compared with the perturbative correlator (see also 
figure~\ref{fig:fgluon}). The center field correlator, 
however, contributes to the operator product corrections. Hence, 
the center vortex ``background'' field could serve a natural explanation for 
the condensates entering the operator product expansion. 
Finally note that $ F^\mathrm{cen}(p^2) $ is enhanced in the low momentum 
regime, where it has the same shape as the 
form factor in Minimal Landau gauge (see figure~\ref{fig:fgluon}).

\vskip 0.3cm  
Let us finally consider the deconfinement phase transition at 
high temperatures. This transition is well understood in the vortex picture 
where it appears as vortex depercolation
transition~\cite{Langfeld:1998cz,Engelhardt:1999fd,Langfeld:2003zi}. 
Since the transition is of 2nd order, it is accompanied 
by the occurrence of a massless excitation. However, it is known that 
neither the gluonic mass gap nor the color singlet states hardly change
significantly and thus cannot be identified with the 
excitation that becomes massless at the transition. 
Because of the success of the vortex picture in 
describing this transition, one might suspect that the center field 
correlator contains the desired information. We have therefore studied 
the correlator $\sum _{\vec{x}} c(t, \vec{x})$ as function of 
(Euclidean) time $t$ for several temperatures. 
Simulations have been carried out using a $30^3 \times 10$ lattice. 
$\beta $ was varied from $1.91$ to $2.69$ to adjust the temperature.
The result is shown in fig. 4. At temperatures $T$ below the 
critical temperature ($T_c$) we find an exponential decrease of the 
correlator $C(t)$ in accordance with the zero temperature 
result~\cite{Polikarpov:2004iv}. For $T>T_c$, the correlation is 
compatible with a power-law indicating a vanishing mass gap 
for the center field propagator. 

\bi
\no
In conclusions, MCG supplemented with $Z_2$-Landau gauge allows for 
a meaningful identification of the $Z_2$ center field and 
$SU(2)/Z_2$ coset parts of the gluon propagator in the continuum 
limit. While the center field contribution dominates the infrared 
behavior of the gluon propagator, it does not contribute to the high 
energy tail of the latter. Furthermore, the center field propagator 
embodies the 
excitations which become massless at the finite temperature $SU(2)$ 
deconfinement phase transition. The ghost form factor of the MCG is 
infrared finite and thus shows that the ``adjoint'' part does not 
play a role for confinement.

%{\it Acknowledgements:}

\end{document}